\documentclass[showpacs,prb,floatfix,twocolumn,amsmath,a4]{revtex4-1}
\usepackage{graphicx}

\begin{document}

\title{Electronic Correlation effects in superconducting picene from ab-initio calculations}

\author{Gianluca Giovannetti$^{1,2}$, Massimo Capone$^{3,1}$}	

\affiliation{$^1$ISC-CNR and Dipartimento di Fisica, Universit\`a di Roma ``La Sapienza'', Piazzale A. Moro 5, 00185, Rome, Italy}
\affiliation{$^2$Institute for Theoretical Solid State Physics, IFW Dresden, 01171 Dresden, Germany}
\affiliation{$^3$Democritos National Simulation Center, CNR-IOM and Scuola Internazionale Superiore di Studi Avanzati (SISSA), Via Bonomea 265, Trieste, Italy}
\date{\today}

\begin{abstract}
We show, by means of ab-initio calculations, that electron-electron correlations play an important  role in potassium-doped picene ($K_x$-picene), recently characterized as a superconductor with $T_c = 18K$. The inclusion of exchange interactions by means of hybrid functionals reproduces the correct gap for the undoped compound and predicts an antiferromagnetic state for $x=3$, where superconductivity has been observed. These calculations, which do not require to assume a value for the interaction strength, indirectly suggest that these materials should have a sizeable ratio between the effective Coulomb repulsion $U$ and the bandwidth. This is fully compatible with simple estimates of this ratio. Using these values of $U$ in a simple effective Hubbard model, an antiferromagnetic state is indeed stabilized. Our results highlight the similarity between potassium-doped picene and alkali-doped fulleride superconductors.
\end{abstract}


\maketitle 

\section{introduction}
The unanticipated discovery of superconductivity with critical temperature $T_c = 18K$ in an aromatic compound like potassium-doped picene\cite{Mitsuhashi} opens a new path to superconductivity in organic materials. Looking for an answer to the most natural question, namely ``What is the origin of the electron pairing?",  we are tempted to consider only two mutually exclusive options:  On one side standard electron-phonon superconductors which are expected to have low critical temperatures, on the other ``anomalous" superconductors, dominated by electron-electron correlation effects. In this latter case pairing is assumed to have electronic origin while the critical temperature may even exceed 100K. 

A challenge to the above distinction comes from a family of organic superconductors, the alkali-metal-doped fullerides. There is wide consensus that pairing is mediated by intramolecular phonons in these materials\cite{gunnarssonreview}. Nonetheless Cs$_3$C$_{60}$, which has one of the largest intramolecular distances in the family, is an antiferromagnetic insulator which turns into a 38K superconductor under pressure\cite{Ganin}. This evidence confirms a previous proposal that correlations are indeed important and helpful for superconductivity in fullerides\cite{capone,capone2,schiro} despite the phononic nature of the pairing "glue".
In this context the discovery of relatively-high $T_c$ in a different organic compound doped with potassium is exciting and brings, among many others, the question about the role of electron-electron correlations. This paper is devoted to a first investigation of the fingerprints of electronic correlations in $K_x$-picene using a wide range of density-functional theory (DFT) methods, including in particular hybrid functionals designed to include some degree of exchange correlations.

Picene is an aromatic molecule (C$_{22}$H$_{14}$) composed by five benzene rings arranged in an ``armchair'' structure which determines a high chemical stability reflected in the band gap of 3.3 eV. Superconductivity has been discovered in solid picene with a monoclinic structure when the band insulating stoichiometric compound is doped by alkali atoms. Indicating with K$_x$-picene a doped solid with $x$ potassium atoms per picene molecule, superconductivity has been observed in the region $ 2.6 < x < 3.3$\cite{Mitsuhashi}. The first ab-initio study of $K_x$-picene has been presented in \cite{Arita}, where the electronic structure and Fermi surface have been obtained within the Local Density Approximation (LDA). In this work we extend this analysis using DFT-based methods aiming to include correlation effects. Namely we compare 
Perdew-Burke-Ernzerhof (PBE)\cite{PBE} implementation of generalized gradient corrections with LDA+U\cite{ldau} calculations for a tight-binding model derived from PBE, and we use Heyd-Scuseria-Ernzerhof (HSE)\cite{hse} and B3LYP\cite{b3lyp} hybrid functionals including different degrees of many-body exchange. 

The paper is organized as follows: Section II presents results for undoped solid picene, while in Section III we turn to describe the doped solid picene estimating a sizable on-site electron-electron interaction, mapping the doped solid picene into an Hubbard model, and then discussing the emergence of an AFM solution.

\section{Bandstructure of undoped picene}

\subsection{Single Picene Molecule}
Given the molecular nature of solid picene, it is useful to recall the basis aspects of the molecular electronic structure, which is expected is the main building block of the bandstructure of the solid. Calculations for an isolated molecule are based on the GAMESS package\cite{GAMESS} and use PBE, PBE0 and B3LYP functionals choosing TZV basis set. A PBE calculation fixing the molecule in its equilibrium geometry in the solid gives a HOMO-LUMO gap of 2.9 eV as reported in \cite{Mitsuhashi}. The gap increases to 4.2 (4.4) eV within B3LYP (PBE0), suggesting an important role of intramolecular exchange interactions. 
Since our focus is on the electron-doped system, the structure of the unoccupied states is of particular importance, and it is characterized by two closely lying states (LUMO and LUMO+1). The difference between these levels is 0.19 (0.20) eV using PBE (B3LYP) and it can shrink down to 75 meV if the geometry of the molecule is optimized, leading to a flatter structure which in turn reduces the molecular overcrowding. On the other hand, using the geometry of the doped crystals, the  LUMO and LUMO+1 gap increases to 0.36 eV. 

The potassium electrons will therefore populate the bands arising from the LUMO and LUMO+1, which can host up to four electrons. We briefly anticipate two possible configurations for $x=3$, the potassium concentration around which superconductivity has been observed: 
\begin{itemize}
\item{(i) the bands originating from LUMO and LUMO+1 may remain distinct  with two electrons per molecule filling the LUMO band and the remaining electron going in the LUMO+1 band, which becomes half-filled.}
\item{(ii) The two bands are not well separated, and the three electrons partially populate both bands so that the system behaves more  like a 3/4-filled two-band system.} 
\end{itemize}
While the former scenario would lead essentially to a single band half-filled Hubbard model which is likely to order antiferrogmagnetically, the latter situation is less favorable for antiferromagnetic ordering and it can give rise to a competition between different phases involving magnetic and charge ordering, as shown by the phase diagrams of $3/4$-filled TMTTF and TMTSF organic compounds\cite{tmttf}. Given the small molecular splitting both situations are in principle possible. We will show that, at least in the present DFT calculations, the first scenario seems to be preferred, even if a more accurate treatment of electron-electron correlations may revive the second possibility.

\subsection{Undoped Solid Picene: Generalized Gradient Approximation and Hybrid Functionals}
\begin{figure}
\includegraphics[width=1.0\columnwidth,angle=0]{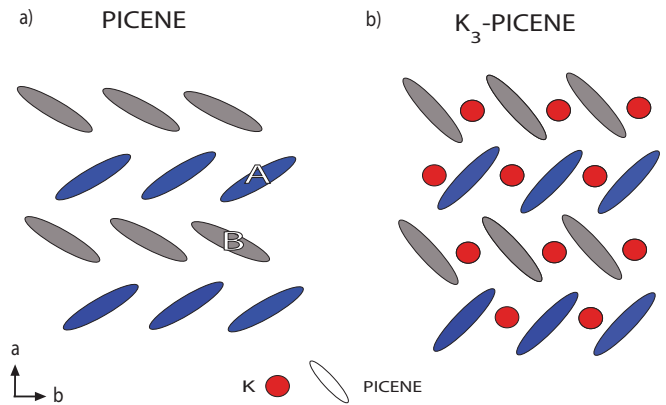}
\caption{(Color online) Schematic arrangement of two inequivalent picene molecules (A and B) along the
  $a$,$b$ planar molecular axes in a) pristine and b) doped molecular crystals.}
\label{fig1}
\end{figure}

We now move to pristine solid picene, that we first study via PBE and the projector augmented wave (PAW) method \cite{paw} using the Vienna \textit{ab-initio} simulation package (VASP)\cite{Kresse}. The valence pseudo-wave-functions were expanded in a plane-wave basis set with a cutoff energy of 400 eV and all the integrations in the Brillouin zone are performed with a Gaussian broadening $\sigma=0.02$ using a sampling grid of 5x7x3 k-points along lattice vectors a,b,c of the unit cell.
The lattice parameters are fixed at the experimental values\cite{picenestructure}. Within the unit cell there are two inequivalent picene molecules (A and B) arranged in herringbone structure in the ab plane with stacking along the c direction (see Fig. \ref{fig1}). 

The bandstructure obtained with PBE confirms the molecular nature of the solid because the relevant bands clearly correspond to the molecular orbitals described above. Our results are in good agreement with Ref. \cite{Arita}, where the relation between the molecular orbitals and the maximally localized Wannier orbitals of the solid has been discussed in details. Since there are two inequivalent picene molecules, the HOMOs of the picene molecules hybridize giving rise  to two bands just below the Fermi level,  which are separated by a band gap of 2.2eV from a manifold of four bands originating from the two LUMOs and two LUMO+1s. As customary, PBE underestimates the experimental value of the gap\cite{gappicenestructure}.
Both the valence and the conduction bands have a bandwidth of around 0.5 eV. The dispersion is enhanced along the a$^*$,b$^*$ directions as expected  from the layered crystal structure. For a similar molecular crystal, Van der Waals interactions may in principle play a role in determining the structure. The reasonable agreement between the experimental structure and the results of our theoretical relaxation suggests that this role is indeed not crucial.

\begin{figure}
\includegraphics[width=1.0\columnwidth,angle=0]{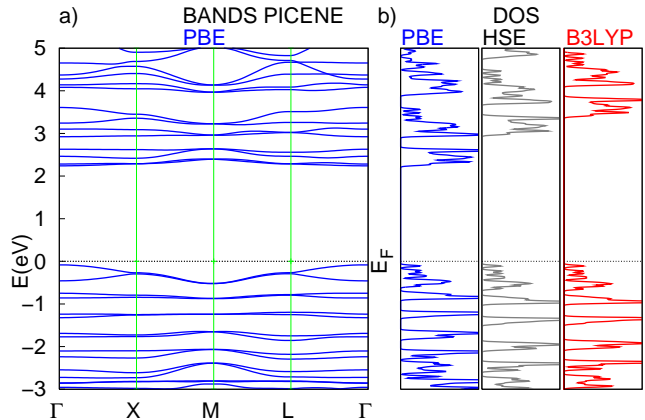}
\caption{(Color online) Results for pristine Picene: Panel (a) presents the PBE bandstructure; panel (b) shows the density of states
  calculated with PBE, HSE, and B3LYP functionals. The energies are plotted along  lines in the Brillouin zone connecting the points $\Gamma=(0,0,0)$,
  $X=(\frac{1}{2},0,0)$, $M=( \frac{1}{2} ,\frac{1}{2},0) $, $L =(\frac{1}{2},0,\frac{1}{2})$. The zero of the energy is at the Fermi-level.} 
\label{fig2}
\end{figure}

To overcome the limitations of PBE (and LDA) and its shortcoming in the determination of the gap, we repeated the same kind of calculations using HSE and B3LYP hybrid functionals, with the same basis set and computational details mentioned above. The two approaches lead to an enhanced gap of 3.0 and 3.25 eV respectively. The inclusion of some degree of short-range exchange interactions (within a Hartree-Fock treatment) is therefore essential to reproduce the experimental gap of 3.3 eV, signaling the important role of intramolecular interactions. An inspection to the density of states obtained via HSE and B3LYP clarifies the role of of the exchange interactions in the undoped system. The bands become significantly narrower with respect to PBE as a consequence of a more localized nature of the carriers, as expected in a correlated system. Correspondingly, the  wavefunctions at each site become even closer to the molecular orbitals, another signature of strong correlations. This effect is particularly strong in the unoccupied bands coming from LUMO and LUMO+1 that the doped electrons are expected to  populate.

\section{Bandstructure of $K_x$-picene}
\subsection{PBE bandstructure}
\begin{figure}
\includegraphics[width=1.0\columnwidth,angle=0]{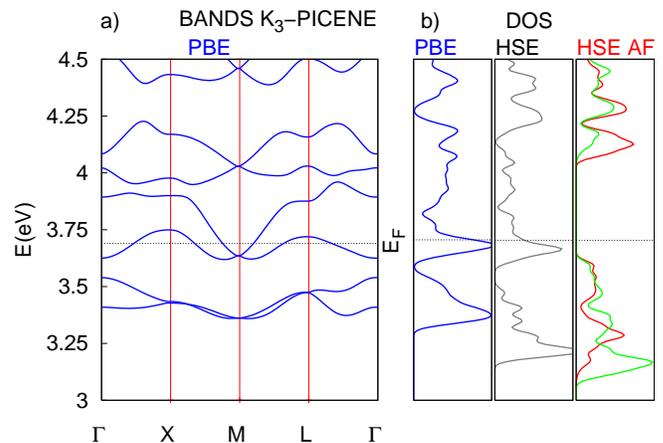}
\caption{(Color online) Results for K$_3$-Picene restricted to the energy region close to the Fermi level: (a) shows the PBE band-structure, while (b) presents the  density of states  calculated with PBE (non magnetic), compared with HSE both in nonmagnetic and AFM state (b). The energies are plotted along  lines in the Brillouin zone connecting the points $\Gamma=(0,0,0)$,  $X=(\frac{1}{2},0,0)$, $M=( \frac{1}{2} ,\frac{1}{2},0) $, $L =(\frac{1}{2},0,\frac{1}{2})$. The zero of energy is at the Fermi-level.}
\label{fig3}
\end{figure}

We finally turn to $K_x$-picene. The experimental results already show that a rigid band picture does not hold, as shown by the difference in the  measured lattice parameters of undoped and K$_{2.9}$-picene, which  suggests that the dopants are not intercalated in the interlayer region and that they affect more deeply the bandstructure, as already discussed in Ref. \cite{Arita,Mitsuhashi}.  We built a unit cell with the lattice parameters measured for K$_{2.9}$-picene and we inserted  3 potassium dopants per picene molecule in the ab plane as suggested in Ref. \cite{Arita,Mitsuhashi} (see Fig. \ref{fig1}). We also repeated within PBE an optimization of the potassium positions starting from different configuration obtaining results essentially equivalent to LDA optimization\cite{Arita}.
The dopant orbitals  are  higher in energy and very weakly hybridized  with the molecular states close to the Fermi level. Therefore their outer s electrons are donated to the picene bands. Indeed the four bands originating from the LUMO and LUMO+1 of the two inequivalent molecules are filled by the six electrons donated by  potassium atoms (three K per picene molecule). Within PBE  (see Fig. \ref{fig3}a) the LUMO and LUMO+1 bands are slightly separated so that the two low-lying bands are completely filled, while the two LUMO+1 bands host an average of one electron per site per band, or, in other words, they are half-filled. The bandwidth $W$ is of around 0.8 eV for all the relevant bands. The above described electronic structure, with one electron per picene molecule in the highest partially occupied band, can give rise to magnetic ordering of the spins of the electrons in the LUMO+1 bands.Within PBE we have stabilized both a ferromagnetic solution and an antiferromagnetic (AFM) solution in which A and B molecules have respectively parallel and opposite spins. Both solutions retain metallic character even if they  present a momentum of around  0.25 $\mu_B$.  Their energy is comparable with the non magnetic state within the accuracy of the calculation.

\subsection{Antiferromagnetic solution within mean-field}
In order to obtain a more reliable picture of the magnetic behavior of the system we need to better account for electron-electron interactions, which are in principle able to give rise to a Mott insulating state with or without magnetic ordering. A simple way is to include the molecular Coulomb repulsion measured by the Hubbard parameter $U$.  It is therefore important to estimate the value of  $U$, which is essentially the energetic cost to doubly occupy a molecule including the screening effects of the other bands\cite{giovannetti}. 
We can compute $U$ for a single molecule with three electrons from the energies of the molecule charged with 2,3 and 4 electrons  as $U=E(4)-2E(3)+E(2)$, where $E(M)$ is the total energy of a molecule charged with $M$ extra electrons, and we will consider $N=3$ in the following.
Such an estimate for an isolated molecule needs to be corrected in order to include the screening effects  in the solid.  A first estimate can be obtained by considering the effect of  the polarization of a charged molecule \cite{bandgap} placed inside a cavity of an homogeneous dielectric medium characterized by a dielectric  constant $\epsilon$. Using typical values for organic molecular crystals $\epsilon =$ 3,4, 5 and 6, we obtain respectively  U $=1.67, 1.43, 1.29, 1.20$ eV\cite{notaUrelaxation}. Even if this estimate can not be regarded as quantitative, given the bandwidth of $W\simeq 0.8 eV$ we obtain a large value of the ratio $U/W$, very close to similar estimates for doped C$_{60}$, where the signatures of the strong repulsion have been experimentally observed\cite{Ganin}. It is therefore important to explore the consequences of electronic correlations also in picene-based solids. A first approach is to introduce the Hubbard repulsion at a mean-field level using the DFT bands as the bare bandstructure. Therefore we built a  tight-binding representation of the bandstructure using Wannier90\cite{wannier90} to compute the maximally localized Wannier orbitals starting from the PBE bandstructure obtained using Quantum Espresso\cite{qe}. Within the tight-binding representation the system is represented as a lattice model, in which the lattice sites coincide with the picene molecules. Indicating with $t^{\alpha\beta}_{ij}$ the hopping amplitude between sites i and j and bands $\alpha$ and$\beta$ and with $\epsilon^{\alpha}_i$ the local energy for an electron in the band $\alpha$ on site i, our tight-binding Hamiltonian reads
\begin{equation}
H = \sum_{i\alpha\sigma} \epsilon^{\alpha}_i c^{\dagger}_{i\alpha\sigma} c_{i\alpha\sigma} + \sum_{ij\alpha\beta} t^{\alpha\beta} (c^{\dagger}_{i\alpha\sigma}c_{j\beta\sigma} + h.c.) + \sum_i \frac{U}{2} n_i^2,
\end{equation}
where $c_{i\alpha\sigma}$ and $c^{\dagger}_{i\alpha\sigma}$ are creation and annihilation operators for electrons of spin $\sigma$ on site i and orbital $\alpha$, and $n_i  = \sum_{\alpha\sigma} c^{\dagger}_{i\alpha\sigma}c_{i\alpha\sigma}$ is the total density on site i. In this preliminary study we limit ourselves to onsite (intramolecular) interactions and we also do not consider exchange interactions between LUMO and LUMO+1. $U$ controls both interband and intraband interactions. In order to implement our mean-field strategy we decouple the interaction term in the particle-hole channel, considering all possible instabilities (charge and magnetic), finding that an antiferromagnetic (AFM) solution with order parameter $m = \sum_i (S^z_{iA} - S^z_{iB})$ ($S^z$ being the z component of the spin on the two sublattices) is clearly favored for our estimated values of $U$.
 Even for the smallest estimate given above (U=1.2 eV), the system becomes AFM (the A molecules having opposite magnetization with respect to B molecules) with a large magnetic moment of 0.96 $\mu_B$ corresponding to (almost) a single spin per picene molecule.  We notice in passing that a tendency towards a charge-ordered solution has been observed within this mean-field approach, even if this solution is never stable. Such a tendency is intriguing because charge-ordering may be expected in 3/4-filled systems in the presence of nearest-neighbor interactions, which are not included here. Non-local interactions may stabilize this phase, reviving the scenario in which the four bands are democratically occupied by the six electrons leading to 3/4-filling.

\subsection{Antiferromagnetic Solution with hybrid functionals}
At a mean-field level, we stabilized an AFM state similar to the parent compound of copper oxide superconductors or to the ground state of Cs$_3$C$_{60}$. One can surmise that also in $K$-picene superconductivity may appear close or compete with an AFM state. It is however well known that the mean-field approach emphasizes the tendency towards antiferromagnetism. Therefore we tested this result using HSE and B3LYP hybrid functionals, also in light of their success in determining the gap of the parent compound that we discussed above. Furthermore, these approach does not require an estimate of $U$, and represent an independent test of the importance of electron-electron correlations. In Fig. \ref{fig3} we show results for HSE, which is believed to perform better in metallic systems  because the long-range tail of the Coulomb kernel is screened \cite{Paier,Batista,Stroppahybrids}, even if the results obtained with B3LYP are fully compatible.
Using both approaches an AFM solution with a reduced magnetic moment of 0.4 $\mu_B$ at each molecular site is indeed stabilized. Within HSE the AFM state has a finite gap of 0.4 eV (see Fig. \ref{fig3}b),  and the gain in energy respect to the non magnetic state is 0.16 eV per unit cell which is at the limit of the accuracy of our calculation.  B3LYP produces an AFM state with the same magnetic moment and slightly larger energy gain (0.22 eV) and gap (0.58 eV). 

Therefore, using the same approaches that correctly reproduce the gap of the parent undoped compound, 
the trivalent doped system appears to be a low-spin AFM, in which less than one electron per picene molecule contributes to magnetism. This result seems to be close to what we obtained by means of static mean-field, with a half-filled conduction band, and the smaller value of the magnetization can be ascribed either to an overestimated $U$ in our mean-field calculation, or to the effect of quantum fluctuations. Ab-initio estimates of $U$ (and possibly of nearest neighbor repulsion and interorbital exchange) will be very helpful to understand this discrepancy, as well as more accurate treatments of short-ranged correlations, such as Dynamical Mean-Field Theory.

\section{conclusions}
In this manuscript, motivated by experiments showing that for $2.7 < x< 3.3$ $K_x$-picene displays superconductivity with $T_c = 18K$, 
we have investigated the bandstructure of this compound with a special focus on the effects of electron-electron correlations. In agreement with previous DFT calculations, we find that the potassium s-electrons are donated to four bands arising from LUMO and LUMO+1 of two inequivalent picene molecules. Six electrons are donated in the case of $K_3$-picene. This gives rise, within our DFT methods, to four completely filled bands arising from the LUMOs while the two bands arising from the LUMO+1s are overall half-filled. Estimates of the correlation strength pose these materials on the correlated side, suggesting that magnetic ordering can be expected. While the tendency toward magnetism is very weak in PBE, the inclusion of correlations at the mean-field level gives a strong tendency towards antiferromagnetism, which is confirmed by the accurate hybrid functional HSE (and also by B3LYP), even if the magnetic moment is substantially reduced. The antiferromagnetic ordering seems quite robust, and it should not depend on the details of the crystal structure, which is not known experimentally for the doped system.

This finding would put $K_x$-picene in a wide class of superconductors in which AFM and superconductivity coexist and/or compete in the same phase diagram. On the other hand, a very recent photoemission study\cite{pes} suggests that a sizable intramolecular electron-phonon coupling is present in $K_x$-picene with $x \simeq 1$, and a role of the same kind of phonons has been proposed in a recent DFT analysis\cite{dft_new}. This is particularly interesting in light of the analogy with fullerides. In these compounds an intra-molecular electron-phonon coupling is indeed believed to provide pairing\cite{gunnarssonreview}, and it has been shown that a similar coupling can indeed be favored by strong Coulomb interaction\cite{capone} because it does not involve charge fluctuations on a molecule, but rather it couples to the internal degrees of freedom, which are not frozen by Coulomb interactions. It appears therefore that this new molecular superconductors shares many similarities with alkali-metal doped fullerides. In this light the AFM state that we find in the present study would be competing with a phonon-driven (and correlation assisted) groundstate, leading to a first-order transition as a function of the cell volume, in close analogy to what observed in Cs$_3$C$_{60}$\cite{Ganin}.

After completion of the manuscript, we became aware of a related work by Kim {\it et al.}, arXiv:1011.2712, in which an AFM state is found also using PBE if the volume per molecule is increased by 5\%. This result is compatible with ours, and confirms the important role of electron-electron correlations in $K_x$-picene.

\section*{Acknowledgments}
We acknowledge financial support by the European Research Council under FP7/ERC Starting Independent Research Grant ``SUPERBAD" (Grant Agreement n. 240524). 
Computational support and computing time from CINECA Supercomputing Center (Bologna) are warmly acknowledged.

The authors thank R. Arita,  A. Stroppa and D. Varsano for fruitful and stimulating discussions.


%
%
%
%

%
%
%



\begin{thebibliography}{100}
\bibitem{Mitsuhashi} R. Mitsuhashi, Y. Suzuki, Y. Yamanari, H. Mitamura, T. Kambe, N. Ikeda, H. Okamoto, A. Fujiwara, M. Yamaji, N. Kawasaki, Y. Maniwa, and Y. Kubozono, Nature {\bf 464}, 76-79 (2010).

\bibitem{gunnarssonreview} O. Gunnarsson, Rev. Mod. Phys. {\bf 69}, 575 (1997). 

\bibitem{Ganin} A.~Y~.Ganin, Y.~Takabashi, Y.~Z.~ Khimyak, S.~Margadonna, A.~Tamai, M.~J.~Rosseinsky, and K.~Prassides, Nature Materials {\bf 7}, 367 (2008);  

Y.~Takabayashi, A. Y. Ganin, P. Jeglic, D. Arcon, T. Takano, Y. Iwasa, Y. Ohishi, M. Takata, N. Takeshita, K. Prassides, and M. J. Rosseinsky, Science {\bf 323}, 1585 (2009).

\bibitem{capone} M. Capone, M. Fabrizio, C. Castellani, and E. Tosatti, Science {\bf 296}, 2364 (2002).

\bibitem{capone2} M. Capone, M. Fabrizio, C. Castellani, and E. Tosatti, Phys. Rev. Lett.  {\bf 93}, 047001 (2004); Rev. Mod. Phys. {\bf 81}, 943 (2009).

\bibitem{schiro} M. Schir\`o, M. Capone, M. Fabrizio, and C. Castellani, Phys. Rev. B {\bf 77}, 104522 (2008).

\bibitem{Arita} T. Kosugi, T. Miyake, S. Ishibashi, R. Arita and H. Aoki, J. Phys. Soc. Jpn. Vol. {\bf 78},  113704 (2009).

\bibitem{PBE} J.P. Perdew, K. Burke, and M. Ernzerhof, Phys. Rev. Lett. {\bf 77}, 3865 (1996).

\bibitem{ldau} V. I. Anisimov, F. Aryasetiawan, and A.I. Lichtenstein, J. Phys.: Condens. Matter {\bf 9}, 767 (1997);

\bibitem{hse}  J.Heyd, G. E. Scuseria, and M. Ernzerhof, J. Chem. Phys. {\bf 118}, 8207 (2003); J. Heyd and G. E. Scuseria, J. Chem. Phys. 124, 219906 (2006).
 

\bibitem{b3lyp} C.~Lee, W.~Yang, and R.~G.~Parr, Phys. Rev. B {\bf 37}, 785 (1988)

\bibitem{GAMESS} M.W. Schmidt, K. K. Baldridge, J. A. Boatz, S. T. Elbert, M. S. Gordon, J. H. Jensen, S.
Koseki, N. Matsunaga, K. A. Nguyen, S. Su, T. L. Windus, M. Dupuis, and J. A.
Montgomery, J. Comput. Chem {\bf 14}, 1347 (1983).

\bibitem{tmttf} S. Brazovskii, in Physics of Organic Superconductors and Conductors, A.G. Lebed, Editor, Springer
Series in Materials Sciences (2008).

\bibitem{paw} G. Kresse and D. Joubert, Phys. Rev. B {\bf 59}, 1758 (1999).

\bibitem{Kresse} G. Kresse and J. Furthmuller, Phys. Rev. B {\bf 54}, 11 169 (1996); Comput. Mater. Sci. {\bf 6}, 15 (1996).

\bibitem{picenestructure} A. De, R. Ghosh, S. Roychowdhury, and P. Roychowdhury, Acta Crystallogr. C {\bf 41} 907 (1985).

\bibitem{gappicenestructure} H. Okamoto, N. Kawasaki, Y. Kaji, Y. Kubozono, A. Fujiwara, and M. Yamaji, J. Am. Chem. Soc {\bf 130}, 10470 (2008).

\bibitem{giovannetti} M.F. Craciun, , G. Giovannetti, S. Rogge, G. Brocks, A. F. Morpurgo, J. van den Brink, Phys. Rev. B {\bf 79}, 125116 (2009);
G. Giovannetti, G. Brocks, and J. van den Brink, Phys. Rev. B {\bf 77}, 035133 (2008);  G. Brocks, J. van den Brink, and A. F. Morpurgo, Phys. Rev. Lett. {\bf 93}, 146405 (2004).

\bibitem{bandgap} To evaluate the polarization energy we use the SS(V)PE model, see D.M. Chipman, J. Chem. Phys. {\bf 112}, 558 (2000).

\bibitem{notaUrelaxation} If the picene molecule is relaxed in the charged state, the effective $U$ is reduced of around 0.15 eV.

\bibitem{wannier90} A. A. Mostofi, J. R. Yates, Y.-S. Lee, I. Souza, D. Vanderbilt and N. Marzari, Comput. Phys. Commun. {\bf 178}, 685 (2008)

\bibitem{qe} P. Giannozzi, S. Baroni, N. Bonini, M. Calandra, R. Car, C. Cavazzoni, D. Ceresoli, G. L. Chiarotti, M. Cococcioni, I. Dabo, A. Dal Corso, S. Fabris, G. Fratesi, S. de Gironcoli, R. Gebauer, U. Gerstmann, C. Gougoussis, A. Kokalj, M. Lazzeri, L. Martin-Samos, N. Marzari, F. Mauri, R. Mazzarello, S. Paolini, A. Pasquarello, L. Paulatto, C. Sbraccia, S. Scandolo, G. Sclauzero, A. P. Seitsonen, A. Smogunov, P. Umari, R. M. Wentzcovitch, J.Phys.: Condens. Matter, {\bf 21}, 395502 (2009) 

\bibitem{Paier} J. Paier, M. Marsman, and G. Kresse, J. Chem. Phys. {\textbf 127}, 024103
  102 (2007).

\bibitem{Batista} E. R. Batista, J. Heyd, R. G. Hennig, B. P. Uberuaga, R. L. Martin, G. E. Scuseria, C. J. Umrigar, and J. W. Wilkins, Phys. Rev. B {\bf 74}, 121102(R) (2006)

\bibitem{Stroppahybrids} M. Marsman, J. Paier, A. Stroppa and G. Kresse, J. Phys.:Condens. Matt. \textbf{20}, 064201 (9) (2008).

\bibitem{pes} H. Okazaki, T. Wakita, T. Muro, Y. Kaji, X. Lee, H. Mitamura, N. Kawasaki, Y. Kubozono, Y. Yamanari, T. Kambe, T. Kato, M. Hirai, Y. Muraoka, T. Yokoya, Phys. Rev. B {\bf 82}, 195114 (2010)

\bibitem{dft_new} P. L. de Andres, A. Guijarro, and J.A. Verge\'es, arXiv:1010.6168




%





\end{thebibliography}
\end{document}